\documentstyle[aps,twocolumn,floats,graphicx]{revtex}

\begin{document}
\title{Doping induced charge redistribution in the high temperature superconductor 
HgBa$_2$CuO$_{4+\delta}$}
\author{P. S\"ule, C. Ambrosch-Draxl, H. Auer and E. Y. Sherman} 
\address{Institut f\"ur Theoretische Physik, Universit\"at Graz, 
Universit\"atsplatz 5, Graz, Austria}
\date{\today}
\maketitle

\begin{abstract}
To understand the link between doping and electronic properties in high temperature 
superconductors, we report first-principles calculations on the oxygen doping
effect for the single layer cuprate HgBa$_2$CuO$_{4+\delta}$. The doping effect is  
studied both by supercell and single cell virtual crystal type approaches.
We find ionic behavior 
of the dopant atom up to an optimal doping concentration of $\delta=0.22$. 
The excess oxygen attracts electrons from the CuO$_2$ plane leading to an increase of 
the hole concentration in this building block. The maximum amount of holes is reached 
when the oxygen $p$-shell is nearly completely filled. As a consequence the density of states at 
the Fermi level exhibits a pronounced maximum at the optimal oxygen content. 
The calculated hole content as a function of doping is in accordance with experimental findings.
%All theoretical findings are in excellent agreement with experimental
%observations. We propose this doping behavior to be a general feature among different 
%families of high temperature superconductors. Maximum $T_c$ is achieved at a
%filled dopant oxygen shell. 
\end{abstract}

\pacs{ }

\narrowtext

It is commonly believed that the charge carriers responsible for the
superconducting pairing of high $T_c$ cuprates are holes mainly confined to 
the CuO$_2$ layers.  
Similar correlations between $T_c$ and and $n_s/m^*$, with $n_s$ and $m^*$ being the 
charge carrier density and the effective mass, respectively, do not only include high 
$T_c$ materials, but also other families of superconductors \cite{Uemura}. 
A universal relationship between $T_c/T_c^{max}$ and the hole content $p$, 
exhibiting a common feature among the $p$-type high-$T_c$ superconductors, 
was presented by Zhang {\em et al.}\cite{Zhang}. This dependence is characterized 
by a plateau, where the variation of $T_c/T_c^{max}$ is independent of the compound 
considered.

In all these materials of interest the actual amount of holes is driven by doping. 
However, for most of the theoretical models describing the superconducting phase 
transition the hole content rather than the doping level is the crucial input parameter. 
Above all, hardly anything is known about the relationship between these two physical 
quantities. Without a detailed knowledge of how doping influences the number of
carriers in the normal state, also the superconducting properties will lack a
profound understanding. In view of being able to tune the amount of 
carriers and thereby the superconducting transition temperature a full
clarification of how doping affects the electronic structure is not only highly 
interesting, but even inevitable. In this context the following questions are raised: 
(i) Where does the ecxess charge go upon doping?
(ii) How does doping influence the carrier concentration in the copper-oxygen planes?
(iii) What limits the amount of holes in these building blocks?
(iv) How does the density of states (DOS) behave as a function of doping?

In order to address these topics, we have carried out electronic structure
calculations based on density functional theory (DFT). For that purpose we have
chosen the simplest representative of the Hg-based high $T_c$ compounds, the CuO$_2$
single-layer HgBa$_2$CuO$_{4+\delta}$ (Hg1201).

This family of layered structures with the formula unit of 
HgBa$_2$Ca$_{n-1}$Cu$_{n}$O$_{2n+2+\delta}$ \cite{Putilin} represents the class of 
superconducting materials with the highest transition temperature so far. It 
reaches a value of 136 K for $n=3$ which can be even enhanced by applying 
pressure \cite{Gao}. Upon adding some excess oxygen, $T_c$ can be 
dramatically increased \cite{Putilin}. For Hg1201, a systematic study over a wider 
range of doping ($0.03 \leq \delta \leq 0.4$) \cite{Xiong} revealed a parabolic-like 
$T_c$ dependence on the oxygen content $\delta$ exhibiting a maximum at 
$\delta_{opt} \simeq 0.22$. 
At this {\it optimal} doping concentration $T_c$ is 97 K for ambient pressure, 
and climbs up to 118 K at 24 GPa, which is the highest $T_c$ of any single-layer
cuprate. Although the mercury based high $T_c$ compounds are not explicitely included 
in the experimental analysis described above \cite{Uemura,Zhang} one can expect them 
to fit into this framework.

Theoretically, doping effects in this class of materials have been studied for a 
high doping level of $\delta=1/2$ only\cite{Singh_93,Singh_94} revealing strong 
covalent bonding in the basal plane with large overlap of the Hg and the dopant
oxygen orbitals. This oxygen concentration, however, has never been achieved 
experimentally. Therefore a detailed study of the much more important low doping 
regime is highly desirable. In our investigations special emphasis is put to 
the doping induced redistribution of the hole concentration in the CuO$_2$ layer. 
In addition, the doping effect on the crystal structure is studied. 
In order to make arbitrary doping levels feasible we treat an excess oxygen 
concentration $\delta$ by adding the appropriate amount of valence electrons
$6 \times \delta$ (6 electrons for the fully doped cell) to the crystal in a 
single-cell calculation. To guarantee charge neutrality the corresponding positive 
charge was placed in a sphere at the doping oxygen site, i.e. in the middle of the 
Hg squares in the basal plane. As we will see below, this virtual crystal approach 
is justified by the ionic behavior of the dopant oxygen. We have cross-checked our 
results by supercell calculations for doping levels of $\delta=1/6$ and $\delta=1/4$,
and in addition for $\delta=1/2$. For the systematic study of the doping-dependent 
electronic structure, the crystalline data as a function of doping were taken from 
Ref. \onlinecite{Huang}. 
In a next step, we also investigate the effect of doping on the 
lattice parameters and atomic positions by total-energy and atomic-force calculations. 

All calculations are carried out within the full-potential linearized augmented 
plane-wave (LAPW) method\cite{Andersen75} utilizing the WIEN97 code \cite{WIEN97}. 
Exchange and correlation effects are treated within the local density approximation 
(LDA). The Brillouin zone integrations are carried out on a $20 \times 20 \times 8$ 
{\bf k}-point mesh (consisting of $220$ points in the irreducible wedge) applying a 
Gaussian smearing of $0.002$ mRy. Our basis set includes approximately 1500 LAPW's 
supplemented by local orbitals for the low-lying semicore-states Hg--$5p$, Cu--$3p$ 
and O--$2s$. We used atomic sphere radii of $2.0, 2.2, 1.9$, and $1.55$ a.u. for Hg, 
Ba, Cu and O, respectively. The atomic-like basis functions used within the LAPW 
method allow an analysis and the orbital symmetry decomposition of the electronic 
charge within the atomic spheres. For the labelling of the oxygen atoms see Fig. 
\ref{fig_cell}.

As already found by previous calculations carried out for the undoped material 
\cite{Singh_93,Rodriguez}, the Fermi level intersects a single free-electron-like 
two-dimensional half-filled $dp\sigma^*$ band, with its states being of 
Cu($d_{x^2-y^2}$) and O($p_x$) character. Upon doping the effect on the electronic 
structure is twofold. First, the dopant adds states at the Fermi level, where 
these new charge carriers cause a shift of $E_F$. Second, the shift of the Fermi 
level changes the carrier concentration in the CuO$_2$ plane. This can be seen in 
Fig. \ref{fig_holes} where the corresponding losses in the partial charges with 
respect to the undoped system are displayed. The occupation number for the copper 
$d_{x^2-y^2}$ orbital $Q(d_{x^2-y^2})$ is decreased by 0.09 $e$ (from 1.443 $e$ for 
the undoped case to 1.353 $e$ for $\delta_{opt}$), and similarly the O1($p_x$) 
charge $Q(p_{x})$ drops by 0.030 $e$ (from 0.982 to 0.952). Both orbitals clearly 
exhibit their minimum occupation number (maximum amount of holes) as a plateau 
around the experimentally observed optimal doping content. The hole concentration 
within the copper-oxygen planes with respect to the undoped case is obtained by 
$[Q(d_{x^2-y^2})(0) + 2 \times Q(p_x)(0)]$ - $[Q(d_{x^2-y^2})(\delta) + 2 \times 
Q(p_x)(\delta)]$ also displayed in Fig. \ref{fig_holes}. In excellent agreement with 
experiment \cite{Xiong} its optimal value is found to be 0.16 $e$. 
%One should note 
%that the partial charges depend on the atomic sphere radii and therefore the 
%absolute values are somewhat bigger; the maximum hole concentration, however, can be 
%estimated to be less than 0.20 $e$. 

The increase of the hole content can be generally understood in an ionic picture: 
The excess oxygen attracts electrons from the copper-oxygen plane thereby increasing 
the hole concentration in this region. In fact, the O3 states do not show strong 
overlap with the neighboring Hg states in the entire doping regime up to optimal 
doping. This picture is confirmed by supercell calculations for doping 
concentrations of $\delta=1/6$ and $\delta=1/4$ (with one excess oxygen O3 per
six and four single cells, respectively). 
It shows a nearly flat oxygen band at $E_F$ with vanishing Hg--$d$ 
admixture, whereas the Hg--$d$ states are found somewhat below. The total and partial 
densities of states for $\delta=1/4$ are shown in Figure \ref{fig_supercell_DOS}. It exhibits 
covalent bonding of Cu and O1 with peaks at -0.35 eV and a small feature right below 
$E_F$. Covalent bonds are also formed by Hg and the apical oxygen O2, whereas the 
dopant atom O3 has the highest contribution to the DOS at the Fermi level with 
hardly any common feature with mercury. This fact is supported by the partial 
charges in the Hg sphere which are not affected by doping. Only in the overdoped 
region, investigated for $\delta=1$, and for $\delta=1/2$ by a supercell calculation, 
stronger bonds between oxygen and mercury atoms are formed as was also found in the 
three-layer compound for the heavily oxygenated material\cite{Singh_94}.

Let us turn now to the question what the limiting fact for the creation of holes in 
the CuO$_2$ plane is. We propose that the hole concentration reaches its 
maximum when the dopant band is nearly completely occupied, i.e. for a {\it closed O3 shell}. 
Since the excess oxygen attracts electrons from the CuO$_2$ plane, its orbitals will be 
systematically filled up not only by the electrons provided by the oxygen atom itself, 
but also by those coming from the CuO$_2$ unit. In order to quantitatively prove our 
hypothesis, we have repeated our calculations with a muffin tin radius of 2.5. a.u. for 
the artificial O3 atoms with nuclear charge of $Z=6 \delta$. This sphere is big enough
to include the doping charge and therefore allows
the interpretation of the charges inside this sphere as absolute occupation numbers.
We indeed find a nearly closed shell at the experimentally observed optimal doping concentraion
$\delta_{opt} = 0.22$ \cite{Xiong} which limits the amount of holes in the CuO$_2$ plane an
therefore most probably the superconducting transition temperature.
Furthermore, the occupation of the $2p_z$ shell of the doping $O$ atom (full oxygen atom here) is found by 4cell calculation. We used a muffin tin sphere pf 3.0 a.u. in this calculation and an O anion
is found with the negative charge of 1.4 $e$. Further increase of the muffin tin sphere
was not possible due to computational difficulties. Therefore, at this moment it is not easy to estimate
whether a $O^{2-}$ anion is formed or an anion with less then 2.0 $e$ negative charge.
Nevertheless, the presence of anion dopant state at the doping site can be clearly seen on the
basis of LDA calculations.
In the overdoped region covalent bonds between O3 and Hg are formed. By this change the pure
ionic character of the dopant oxygen gets lost and therefore no more carriers are attracted
from the CuO$_2$ plane which results in the broad plateau on Fig. \ref{fig_holes} in 
accordance with the experimental findings \cite{Xiong}.

These findings also show up in the density of states which in the optimal 
doping regime exhibits an almost filled band right below $E_F$. 
The calculated DOS at the Fermi level $N(E_F)$ as displayed in Fig. 
\ref{fig_DOS} has a pronounced maximum at $\delta_{0} = 0.22$. It contains large 
contributions from the CuO$_2$ plane, but also a considerable amount of apical oxygen 
and some O3. All the contributions peak at the optimal doping content. 
%                       
%The change in the O3-DOS at E$_F$ close to optimal doping can be understood in terms 
%of the charge redistribution: The charge carriers attracted from the planes change their
%atomic character from "plane-like" to "O3-like", and thus are most localized
%when the O3 shell is completely filled. 
%
%For higher doping levels the O3 band is pushed down in energy leading to a decrease in $N(E_F)$. 
%In the overdoped region covalent bonds between O3 and Hg are formed. By this change the pure 
%ionic character of the dopant oxygen gets lost and therefore no more carriers are attracted from the
%CuO$_2$ plane.
%
>From the density of states at $E_F$ the electronic specific heat coefficient can be
determined. Assuming a similar behavior for other high-$T_c$ materials like the 
Bi-based compounds, the measured specific heat coefficient $\gamma$ of 
Bi$_2$Sr$_2$CaCu$_2$O$_{8+\delta}$ as a function of doping \cite{Tallon} can be
understood by the pronounced increase of the DOS around optimal doping. 
Moreover, the calculated maximum value of $\gamma$=4.4 mJ/(mol K$^{2}$) 
is in excellent accordance with experimental values for other high $T_{c}$ 
materials at optimal doping \cite{Plakida}. 

As already mentioned, we crosschecked our results with supercell calculations
for $\delta=1/6$ and $\delta=1/4$ in order to ensure that no artifacts may arise due to 
our single-cell approach for the treatment of doping. The partial charges in all the 
atomic spheres agree very well with the corresponding single-cell calculation (within 
one hundredth of an $e$). E.g. for $\delta=1/4$ the copper $Q(d_{x^2-y^2})$ partial charge 
is 1.364 compared to 1.353 in the single-cell approach, the oxygen $Q(p_x)$ charges are 0.949 
and 0.952, respectively. For $\delta=1/6$ the deviations are also below 1\%. 
Excellent agreement is also found for the density of states at the Fermi level, 
as can be seen for both cases in Fig. \ref{fig_DOS}.

Not only the electronic properties, but also the crystal structure is influenced by doping.
The redistribution of bond lengths can again be explained within the ionic picture. 
Since doping does not change the symmetry of the crystal, it only leads to 
displacements of Ba and the apical oxygen in $z$-direction. With 
increasing oxygen content at the doping site, the positive Ba ions are expected to be 
attracted by the oxygen, whereas the negatively charged O2 ions should be repelled. 
Indeed a shift of the Ba position towards the basal plane is observed together with a
displacement of the O2 ions away from this plane. In Fig. \ref{fig_bonds} the 
corresponding $z$-coordinates of both atoms are shown which were calculated by fully 
optimizing the crystalline data for selected doping concentrations: 
For $\delta$=0, 0.08, 0.17, and 0.23 the unit cell volumes, $c/a$ ratios, and atomic 
positions were obtained by total-energy and atomic-force calculations using a similar 
procedure like in Ref. \onlinecite{Kouba_99}. The results are displayed in 
Table \ref{tab_lattice}.
All trends are in agreement with experimental observations \cite{Huang}: 
These are a shrinkage of the unit cell upon doping, a shortening of the lattice
parameter $a$, an increase of $c$ for small doping levels followed by a nearly
linear decrease, an increasing distance between the basal plane and the Ba
layer, as well as a reduction of the distance between the layers of Ba and O2.
In agreement with measured data, the latter two quantities show a less pronounced
dependence up to $\delta=0.08$ and a stronger one for higher oxygen contents.

In summary, we have investigated the relationship between doping and electronic
and structural properties in the high temperature superconductor HgBa$_2$CuO$_{4+\delta}$. 
We find that for doping concentrations up to optimum the dopant does not form 
covalent bonds with the neighboring Hg atoms, but shows purely ionic behavior. 
The doping oxygen adds states close to the Fermi level showing nearly 
pure O3 character. Due to its high electronegativity the excess oxygen attracts 
electrons from the CuO$_2$ plane resulting in a redistribution of charge carriers 
and an increase of the number of holes in this plane. At optimal doping the DOS at the 
Fermi level exhibits a pronounced maximum in accordance with experimental 
specific heat data of other cuprates. The doping-induced charge redistribution 
is accompanied by a shrinkage of the unit cell and displacements of Ba and the 
apical oxygen O2 towards and from the basal plane, respectively. The amount of holes 
created in the CuO$_2$ plane is limited by the fact that at optimal doping a closed oxygen 
shell is reached. We conclude that this behavior is a general feature among different families 
of high $T_c$ compounds.

\medskip
\noindent
{\bf Acknowledgments} \newline
This work was supported by the Austrian Science Fund, project P13430-PHY. 
We appreciate very stimulating discussions with J. O. Sofo.
%acknowledges conversation with J. Pipek and A. Zawadowski. 

\newpage

\begin{table}
\caption[]
{Crystalline data as a function of doping obtained from total-energy and
atomic-force calculations.}
\begin{tabular}{c|ccccc}
 $\delta$ &  $a$ [\AA] & $c$ [\AA] & $\Omega$ [\AA$^3$] & $z$(Ba) [$c$] & $z$(O2)[$c$]\\ \hline

  0.00  &  3.8574  &  9.7228  & 144.6676  &  0.3003  & 0.2014 \\
  0.08  &  3.8192  &  9.8191  & 143.2211  &  0.2965  & 0.2023 \\
  0.17  &  3.8058  &  9.6887  & 140.3283  &  0.2851  & 0.2055 \\
  0.23  &  3.8312  &  9.5604  & 140.3280  &  0.2808  & 0.2074 \\
\end{tabular}
\label{tab_lattice}
\end{table}

\newpage

%\begin{figure}[]
%\begin{center}
%\includegraphics*[height=6.5cm]{hg1201.ps}
%\caption[]{Crystal structure of HgBa$_2$CuO$_{4+\delta}$.}
%\label{fig_cell}
%\end{center}
%\end{figure}

\begin{figure}[hbtp]
\begin{center}
\caption[]{Doping induced changes in the partial charges within the atomic
spheres of Cu (half filled circles) and O1 (open circles) and total amount of holes 
(filled circles) in the CuO$_2$ plane as a function of the doping concentration $\delta$.}
\label{fig_holes}
\end{center}
\end{figure}

%\newpage

\begin{figure}[hbtp]
\begin{center}
\caption[]{Total and site-projected densities of states in states per eV and
unit cell volume $\Omega$ for $\delta=0.25$ obtained by a supercell calculation 
containing one O3 atom per four formula units. (The Fermi level is set to zero.)}
\label{fig_supercell_DOS}
\end{center}
\end{figure}

\newpage

\begin{figure}[]
\begin{center}
\caption[]{Total density of states at $E_F$ per eV and formula unit for
HgBa$_2$CuO$_{4+\delta}$ as a function of doping. The right axis indicates the
electronic linear specific-heat coefficient $\gamma$ in mJ per mol and K$^2$.
The maximum value is found exactly at the experimentally observed optimum doping
concentration. The results for the supercell calculations are indicated by stars.}
\label{fig_DOS}
\end{center}
\end{figure}

\begin{figure}[hbtp]
\begin{center}
\caption[]{Doping induced changes in the atomic positions of Ba and O2 in units
of the lattice parameter $c$ as function of doping.}
\label{fig_bonds}
\end{center}
\end{figure}

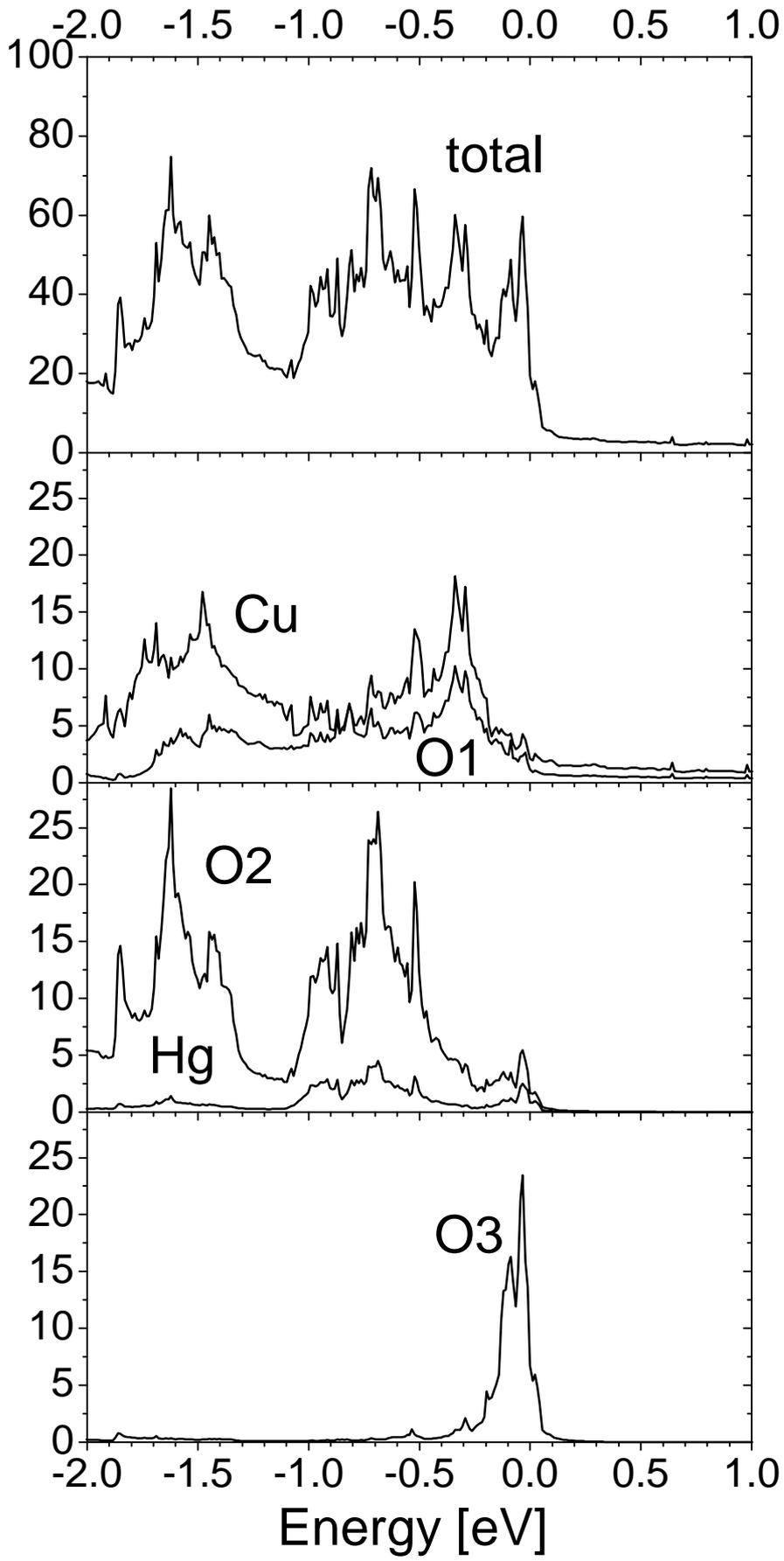
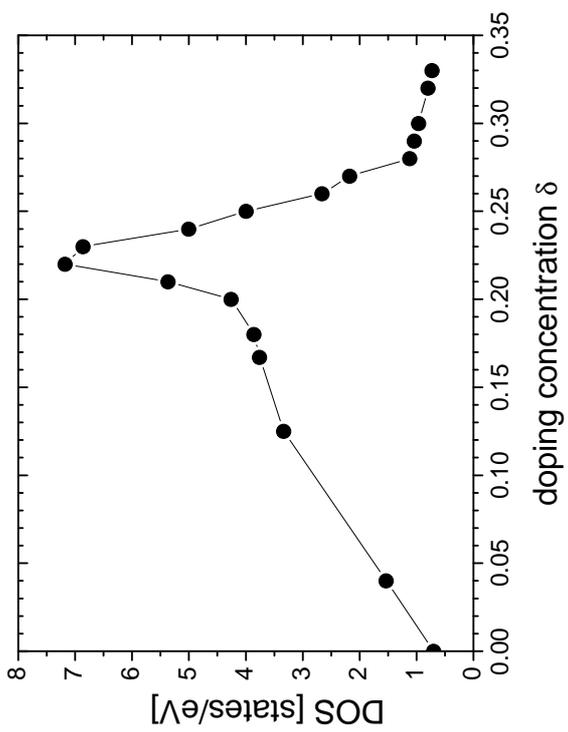
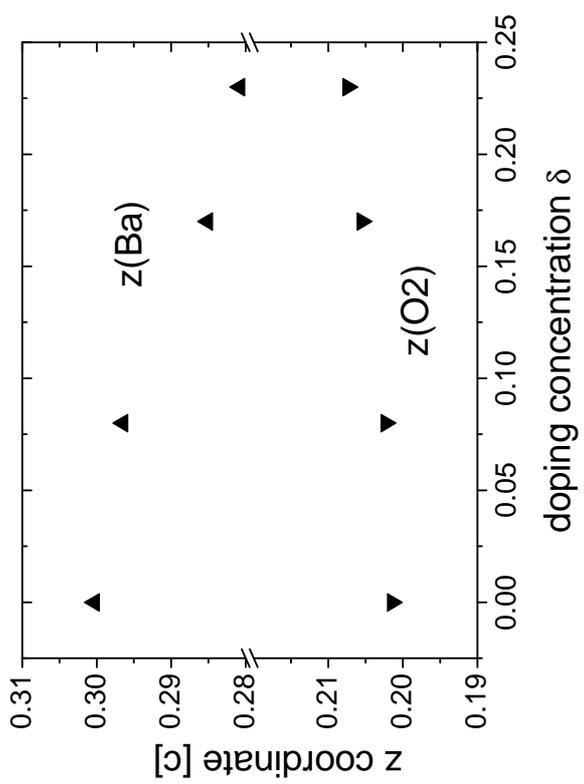
\end{document}